\documentclass[preprint,tightenlines,aps,prd,showpacs]{revtex4}
\usepackage{graphicx}
\usepackage{amsmath}
\usepackage{bm}

\begin{document}

\title{\mbox{}\\[10pt]
Production of the $X(3870)$ at the $\Upsilon(4S)$\\
by the Coalescence of Charm Mesons}

\author{Eric Braaten and Masaoki Kusunoki}
\affiliation{
Physics Department, Ohio State University, Columbus, Ohio 43210}

\date{\today}
\begin{abstract}
If the recently-discovered charmonium state $X(3870)$ is a 
loosely-bound molecule of the charm mesons 
$D^0$ and $\bar D^{*0}$ or $\bar D^0$ and $D^{*0}$,
it can be produced in $e^+ e^-$ annihilation at the $\Upsilon(4S)$
resonance by the coalescence of charm mesons produced 
in the decays of $B^+$ and $B^-$ mesons.
Remarkably, in the case of 2-body decays of the $B$ mesons,
the leading contribution to the 
coalescence probability depends only on hadron masses 
and on the width and branching fractions of the $B$ meson.
As the binding energy $E_b$ of the molecule goes to zero, 
the coalescence probability  scales as $E_b^{1/2} \log(E_b)$.
Unfortunately, the coalescence probability 
is also suppressed by two powers of the ratio
of the width to the mass of the $B$ meson, 
and is therefore many orders of magnitude too small 
to be observed in current experiments at the $B$ factories.
\end{abstract}



\maketitle


The recent unexpected discovery 
of a narrow charmonium resonance near 3.87 GeV 
presents a challenge to our understanding of heavy quarkonium.
The new charmonium state $X(3870)$ was discovered  
by the Belle collaboration in electron-positron 
collisions through the $B$-meson decay $B^\pm  \to K^\pm X$
followed by the decay $X  \to J/\psi \pi^+ \pi^-$ \cite{Choi:2003ue}.
Its mass was measured to be $M_X=3872.0 \pm 0.6 \pm 0.5$ MeV
\cite{Choi:2003ue}.
It is narrow compared to other charmonium states  
above the threshold for decay into $D \bar D$:
the upper bound on the width is $\Gamma_X < 2.3$ MeV.
The discovery has been confirmed by the CDF collaboration 
who observed $X$ through $J/\psi \pi^+ \pi^-$ events in 
proton-antiproton collisions and measured its mass to be
$M_X = 3871.4 \pm 0.7 \pm 0.4$ MeV \cite{Acosta:2003zx}.

The proposed interpretations of the $X(3870)$ 
include a D-wave charmonium state with 
quantum numbers $J^{PC}= 2^{--}$ or $2^{-+}$,
an excited P-wave charmonium state 
with $J^{PC}= 1^{++}$ or $1^{+-}$,
a ``hybrid charmonium'' state in which a gluonic mode 
has been excited, and a $D^0 \bar D^{*0}$  
or $\bar D^0 D^{*0}$ molecule 
\cite{Tornqvist:2003na,Close:2003sg,Pakvasa:2003ea,Voloshin:2003nt,Yuan:2003yz,Wong:2003xk,Braaten:2003he,Barnes:2003vb,Swanson:2003tb,Eichten:2004uh}. 
This last possibility 
is motivated by the fact that the $X(3870)$ is
extremely close to the threshold $3871.2 \pm 0.7$ MeV for decay
into the charmed mesons $D^0 $ and $\bar D^{*0}$.  

If the $X(3870)$  is an S-wave $D^0 \bar D^{*0} /\bar D^0 D^{*0}$ 
molecule, its binding energy is smaller than any other 
hadron that can be interpreted as a 2-body bound state of hadrons.
For two hadrons whose 
low-energy interactions are  mediated by 
pion exchange, the natural low-energy scale for the binding energy of 
a molecule is $m_\pi^2/(2 m)$,
where $m$ is the reduced mass of the two hadrons.  
The natural low-energy scale for a  $D^0 \bar D^{*0}$ molecule 
is about 10 MeV.
The measurements of the mass
of the $X(3870)$ imply that its binding energy
(which is positive by definition)
is $E_b = -0.5 \pm 0.9$ MeV.  
Thus $E_b$ is likely to be less than 0.4 MeV,
which is much smaller than the natural low-energy scale.

The tiny binding energy of the $X(3870)$ implies 
that the $D^0 \bar D^{*0}$ scattering length $a$ is unnaturally large
compared to the natural scale $1/m_\pi$.  
The molecule therefore has properties that depend on $a$ but are 
insensitive to other details of the interactions of $D^0$
and $\bar D^{*0}$, a phenomenon called ``low-energy universality.''  
The universal prediction for the binding energy is 
\begin{eqnarray}
E_b \equiv m_D+m_{D^*} -M_X \simeq {1 \over 2 m_{DD^*} a^2},
\label{B2}
\end{eqnarray}
where $m_{DD^*} = m_{D^0} m_{D^{*0}}/(m_{D^0}+m_{D^{*0}})$
is the reduced mass of the two constituents. 
The universal prediction for the momentum space wavefunction of the 
$D^0 \bar D^{*0}$ or $\bar D^0 D^{*0}$ is 
\begin{eqnarray}
\psi(k) \simeq {(8\pi/a)^{1/2} \over \bm{k}^2+1/a^2} 
\hspace{1.5cm} |\bm{k}|\ll m_\pi,
\label{psi}
\end{eqnarray}
where the normalization is $\int d^3k/(2\pi)^3 |\psi(k)|^2=1$.
This wavefunction has been exploited by Voloshin 
to calculate the momentum distributions for the decays 
$X \to D^0 \bar D^0 \pi^0$ and $X \to D^0 \bar D^0 \gamma$ 
\cite{Voloshin:2003nt}.
The universal prediction for the amplitude for elastic scattering of
$\bar{D}^0$ and $D^{*0}$ in the center-of-momentum frame with total
energy $E$ is
\begin{align}
 {\cal A}[\bar{D}^0D^{*0} \to \bar{D}^0D^{*0}]
 \simeq 
\frac{8\pi m_D m_{D^*}}{m_{DD^*} \left(-1/a-i|\bm{k}| \right)}
\hspace{1.5cm} |\bm{k}|\ll m_\pi,
\label{eq:scat}
\end{align}
where $|\bm{k}|=[2m_{DD^*}(E - m_D - m_{D^*})]^{1/2}$.
Other consequences of low-energy universality have been discussed 
in Ref.~\cite{Braaten:2003he}.
One consequence is that as the scattering length $a$
increases, the probabilities for components of 
the wavefunction other than
$D^0 \bar D^{*0}$ or $\bar D^0 D^{*0}$ decrease as $1/a$.
In the limit $a \to \infty$,
it becomes a pure molecular state. 
If it has charge conjugation
$\cal C = \pm$, the state is a superposition
$(|D^0 \bar D^{*0}\rangle \pm |\bar D^0 D^{*0}\rangle )/\sqrt{2}$.

One of the challenges for the interpretations of the $X(3870)$
as a $D^0\bar{D}^{*0}/\bar{D}^0D^{*0}$ molecule is to understand 
its production rate. The large $D^0\bar{D}^{*0}$ scattering length
implies that a $D^0$ and $\bar{D}^{*0}$ with relative
momentum small compared to $m_\pi$ have very strong interactions.
One way to produce $X$ is therefore to produce $\bar{D}^0$ and $D^{*0}$
with small relative momentum which then undergo a final-state
interaction that binds them to form $X$. An example of such a process
is the decay of $\Upsilon(4S)$ into $B^+B^-$, 
followed by decays of the
$B^+$ and $B^-$ into states containing $\bar{D}^0$ and $D^{*0}$,
respectively. There is a small probability that the $\bar{D}^0$ and
$D^{*0}$ will be produced with sufficiently small relative momentum for
them to coalesce into $X$. In this paper, we calculate the leading
contribution to the coalescence probability in the case of 2-body
decays of the $B^+$ and $B^-$.
We show that the coalescence probability scales as $E_b^{1/2}\log E_b$
as the binding energy of $X$ goes to 0.
Remarkably, the coefficient of $E_b^{1/2}\log E_b$ depends only on
hadron masses and on the width  and branching fractions of the $B$ meson.

We consider the decay $\Upsilon(4S)\to Xhh'$, where $h$ and $h'$ are light
hadrons. 
This process can proceed via the decay $\Upsilon(4S)\to B^+B^-$, 
followed by the 2-body decays 
$B^+\to\bar{D}^0 h$ and $B^-\to D^{*0}h'$, 
followed by the coalescence $\bar{D}^0 D^{*0} \to X$. This process can
also proceed through a second pathway consisting of the 2-body
decays $B^+\to\bar{D}^{*0} h$ and $B^-\to\ D^0 h'$ followed by the
coalescence $ \bar{D}^{*0} D^0 \to X$. 
In principle, these two pathways can interfere.
However, we shall see that the momentum configurations of 
$X$, $h$ and $h'$ are completely
determined by the masses of the hadrons 
and they are different for the two pathways.
Thus there is no interference between the amplitudes.

The decay process $\Upsilon(4S)\to Xhh'$ is complicated because there are
many relevant energy and momentum scales and they range over many
orders of magnitude. The mass $M_\Upsilon$ of the $\Upsilon(4S)$ is larger than the
binding energy $E_b$ of $X$ by more than 4 orders of magnitude and the
width $\Gamma_B$ of the $B$ meson is smaller by about $10$ orders of
magnitude. We expect the rate for this decay to go to $0$ in the limit
$E_b\to 0$ (with $E_b \gg\Gamma_B$), because the $\bar D^0$ and $D^{*0}$
must have relative momentum $k$ of order 
$(m_{DD^*} E_b)^{1/2} \approx 1/a$ in order to bind to form $X$ and such
small relative momentum accounts for a decreasing fraction of the total
phase space available to the $\bar D^0$ and $D^{*0}$.
Our calculation shows that there are contributions to the rate that
scale as $E_b^{1/2}$. They include contributions from relative momentum
$k$ ranging from the scale $1/a$ to the scale $m_\pi$. The contributions
from $k\sim 1/a$ are constrained by low-energy universality, and we
expect these to be calculable in terms of the scattering length.
The contributions from $k\sim m_\pi$ necessarily involve the full
complications of low-energy QCD. Fortunately we find that there is a
logarithmic contribution coming from the range $1/a \ll k \ll m_\pi$,
which is also governed by low-energy universality. This logarithmic term
dominates in the limit $E_b\to 0$.
The logarithmic term in the decay rate for $\Upsilon(4S)\to Xhh'$ is calculated
in Appendix A. The branching ratio 
that measures the coalescence probability for 
$\bar D^0 D^{*0} \to X$ or $\bar D^{*0} D^0 \to X$ is 
\begin{align}
 & \frac{\Gamma[\Upsilon(4S)\to Xhh']}
{\Gamma[\Upsilon(4S)\to \bar{D}^0D^{*0}hh'] + 
 \Gamma[\Upsilon(4S)\to \bar{D}^{*0}D^{0}hh']}
\nonumber \\
& \hspace{1.5cm}
=\frac{ 2\pi M_X m_B^8 }
       { m_{DD^*} M_\Upsilon (M_\Upsilon^2-4m_B^2)^{1/2}} \,
       \left(\frac{2E_b}{m_{DD^*}}\right)^{1/2}\,
       \log\left(\frac{m_\pi^2}{2m_{DD^*}E_b}\right) 
       \left(\frac{\Gamma_B}{m_B}\right)^2
\nonumber \\
 & \hspace{1.5cm}
\times \sum   
  {\cal B}[B^+\to \bar D^{0} h] {\cal B}[B^-\to D^{*0} h']\,
     \frac{J(M_\Upsilon, m_B, M_X, m_D, m_{D^*}, m_h, m_{h'})}
       { \lambda^{1/2}(m_B,m_D,m_h) \,  
         \lambda^{1/2}(m_B,m_{D^*},m_{h'})} 
\nonumber \\
 & \hspace{1.5cm} 
 \times
  \left( \sum {\cal B}[B^+\to \bar D^{0} h] 
        {\cal B}[B^-\to D^{*0} h'] \right)^{-1},
\label{eq:ratio}
\end{align}
where $J(M_\Upsilon,\cdots)$ 
is the function of hadron masses given in
(\ref{eq:J}) and the function $\lambda(m_1,m_2,m_3)$ is given after 
(\ref{eq:B}).
The sum in the numerator and the denominator is over two terms, the one
shown and a second term obtained by replacing $\bar D^0$ and $D^{*0}$
by $\bar D^{*0}$ and $D^0$.
Notice that the expression (\ref{eq:ratio}) depends only on hadron masses 
and on the width $\Gamma_B$ and branching fractions of the $B$-meson. 
If $h$ and $h'$ are each others antiparticles
such as $\pi^+$ and $\pi^-$, the
branching fractions cancel between the numerator and denominator.
If we take the binding energy of $X$ to be
$E_b=0.1$ MeV, 
the branching ratios in (\ref{eq:ratio}) for the cases 
$hh' = (\pi^+\pi^-, \rho^+\rho^-, K^+K^-, K^{*+}K^{*-})$
are $(1.1, 1.3, 1.2, 1.3)\times 10^{-24}$.
For any other combination of $h=(\pi^+,\rho^+,K^+, K^{*+})$ and 
$h'=(\pi^-,\rho^-,K^-, K^{*-})$, the branching ratio in (\ref{eq:ratio})
depends on $B^+$ branching fractions but it is in the range from 
$1.2$ to $1.4 \times 10^{-24}$.

We can get a simple expression that can be used to 
estimate the order of magnitude of the branching ratio by 
neglecting the light hadron masses
$m_h$ and $m_h'$, and 
making the approximations
$m_{D^*}-m_D \ll M_X$ and $M_X \ll m_B$.
In this limit, the function $J(p_Q, p'_Q)$ given by (\ref{eq:J}) approaches
\begin{align}
J(p_Q, p'_Q)
\longrightarrow
\frac{\pi M_X}{m_B^2 (M_\Upsilon^2-4m_B^2)^{1/2}}.
\label{eq:Japprox}
\end{align} 
The branching ratio in (\ref{eq:ratio}) then reduces to 
\begin{align}
\frac{\Gamma[\Upsilon(4S) \rightarrow X h h']}
{\Gamma[\Upsilon(4S)  \rightarrow \bar{D}^0 D^{*0} h h']
+  \Gamma[\Upsilon(4S)\to \bar{D}^{*0}D^{0}hh']} &
\nonumber \\
& \hspace{-4cm}
\longrightarrow
 \frac{8\pi^2 m_B^2 M_X}{M_\Upsilon (M_\Upsilon^2-4m_B^2)} 
 \left(\frac{8E_b}{M_X}\right)^{1/2}
 \log\left(\frac{2 m_\pi^2}{M_X E_b}\right)
 \left(\frac{\Gamma_B}{m_B}\right)^2.
\end{align} 
If we take the binding energy to be $0.1$ MeV,
this estimate for the branching ratio is $ 6.3 \times 10^{-25}$,
which is within a factor of 2 of the more accurate results
calculated using (\ref{eq:ratio}). This estimate applies equally well 
if $h$ or $h'$ is replaced by a
multiparticle state of light hadrons or a lepton pair whose invariant
mass is small compared to $m_B$. We conclude that the inclusive branching
fraction for $\Upsilon(4S)\to X(3870)$ via this coalescence mechanism is
about 24 orders of magnitude smaller than the product of the inclusive
branching fractions for $B^+ \to \bar D^0$ and $B^+ \to  \bar D^{*0}$.

We have calculated the leading contribution to the probability for 
charm mesons produced by the decay of  
$\Upsilon(4S)$ to coalesce into $X(3870)$.
Remarkably, this coalescence probability can be expressed completely in
terms of hadron masses and the width and branching fractions of the $B$
meson. Unfortunately
there is a suppression factor of $(\Gamma_B/m_B)^2$ that makes the rate
for $\Upsilon(4S)\to Xhh'$ many orders of magnitude too small to be observed
at the current $B$ factories.

This research was supported in part by the Department of Energy under
grant DE-FG02-91-ER4069.

\appendix

\section {Calculation of the rate for $\bm{\Upsilon(4S) \longrightarrow X h h'}$}
\begin{figure}
\includegraphics{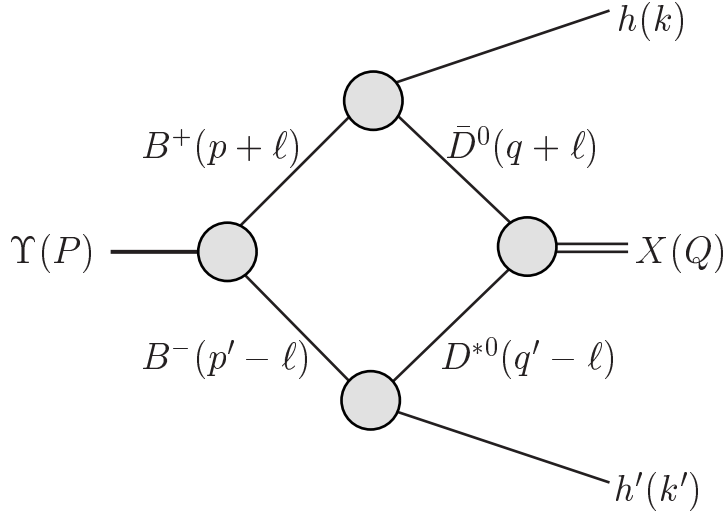}
\caption{Feynman diagram for the amplitude of 
$\Upsilon(4S)\rightarrow X h h'$ via the first pathway. 
\label{fig1}}
\end{figure}

In this appendix, we calculate the rate for the decay 
$\Upsilon(4S)\to X(3870) + h+h'$, where $h$ and $h'$ are light hadrons.
This decay proceeds through two pathways: 
the decay $\Upsilon(4S)\to B^+B^-$ followed by
the 2-body decays $B^+\to \bar{D}^0 h$ and $B^-\to D^{*0} h'$ followed by the
coalescence process $\bar{D}^0 D^{*0}\to X$,
and the pathway obtained by replacing $\bar D^0$ and $D^{*0}$ by $\bar D^{*0}$
and $D^{0}$. As we shall see, the two pathways do not interfere.
The amplitude for the first pathway can be represented by the 1-loop 
Feynman diagram with meson lines shown in Fig.~\ref{fig1}. 
We denote the $\Upsilon(4S)$ simply by $\Upsilon$.
The momenta of $\Upsilon$, $X$, $h$ and $h'$ are 
$P$, $Q$, $k$ and $k'$, respectively. 
The momenta of the virtual $B^+$, $B^-$, $\bar{D}^0$ and $D^{*0}$ mesons
are $p+\ell$, $p'-\ell$, $q+\ell$ and $q'-\ell$, respectively, where 
$\ell$ is the loop momentum.
Momentum conservation requires $P=p+p'$, $p=k+q$, $p'=k'+q'$, and
$q+q'=Q$. We can choose $q$ and $q'$ to be 
\begin{align}
 \begin{split}
  q^{\mu} &= \frac{m_D}{M_X}Q^{\mu}, \\
  q'^{\mu} &= \frac{m_{D^*}}{M_X}Q^{\mu}, 
 \end{split}
\label{eq:qq'}
\end{align} 
so that they are on the mass shells of
the $\bar{D}^0$ and $D^{*0}$:
$q^2=m_D^2$ and $q'^2=m_{D^*}^2$. 
Since the binding energy of $X$ is so tiny, 
the momenta (\ref{eq:qq'}) are consistent with the constraint 
$q+q'= Q$.

The decay rate can be written as 
\begin{align}
\Gamma[\Upsilon  \rightarrow X h h'] =
\frac{1}{2 M_{\Upsilon}} 
\int &
 \frac{d^3k}{(2\pi)^3 2 k_0}
\frac{d^3k'}{(2\pi)^3 2 k'_0}\frac{d^3Q}{(2\pi)^3 2 Q_0}
\nonumber \\ &
\times
(2\pi)^4 \delta^{(4)}(P-k-k'-Q) 
\bigl|
{\cal A}
[\Upsilon \rightarrow X h h']
\bigr|^2.
\end{align} 
The amplitude for the decay through the first pathway is
\begin{align}
{\cal A}_1[\Upsilon \rightarrow X h h']
= \int \frac{d^4\ell}{(2\pi)^4} &\;
{\cal A}[\Upsilon \rightarrow B^+ B^-]
{\cal A}[B^+ \rightarrow \bar{D}^0 h]
{\cal A}[B^- \rightarrow D^{*0} h']
{\cal A}[\bar{D}^0D^{*0} \rightarrow X]
\nonumber \\
& \times
\frac{i}{(p+\ell)^2-m_B^2+ im_B\Gamma_B } \,
\frac{i}{(p'-\ell)^2-m_B^2+ im_B\Gamma_B }
\nonumber \\
& \times
\frac{i}{(q+\ell)^2-m_D^2+ i\epsilon}\,
\frac{i}{(q'-\ell)^2-m_{D^*}^2+ i\epsilon}.
\label{eq:amplitude}
\end{align}
The rate depends crucially on the width $\Gamma_B$ of the $B$ meson,
so the effect of the width must be included in the propagators of
the $B^+$ and $B^-$.

Since the loop integral is dominated by very small momenta, 
we can neglect any momentum dependence of the 
amplitudes $\cal{A}$ for $\Upsilon\to B^+B^-$, $B^+\to\bar{D}^0h$ and
$B^-\to D^{*0} h'$.  
They can be approximated by the amplitudes for the on-shell decays.
For example, the amplitude for $B^+\to \bar{D}^0h$ is determined by
the branching fraction for that decay:
\begin{align}
 {\cal B}[B^+\to \bar{D}^0h]=\frac{1}{16\pi}
\left|{\cal A}[B^+\to \bar{D}^0h]\right|^2 
\frac{\lambda^{1/2}(m_B,m_D,m_h)}{m_B^3 \Gamma_B},
\label{eq:B}
\end{align}
where
$\lambda(x,y,z) = x^4 + y^4 + z^4 -2 (x^2y^2+y^2z^2+z^2x^2)$. 
The amplitude for the coalescence process 
$\bar{D}^0D^{*0} \rightarrow X$ 
can be deduced from the amplitude (\ref{eq:scat}) for 
the scattering process 
$ \bar{D}^0D^{*0} \to \bar{D}^0D^{*0}$. 
This amplitude has a pole in the total energy
$E$ at the mass $M_X=m_D+m_{D^*}-E_b$, where $E_b$ is the binding
energy given by (\ref{B2}). Its behavior near the pole is 
\begin{align}
 {\cal A}[\bar{D}^0D^{*0} \to \bar{D}^0D^{*0}]
\longrightarrow 
 -\frac{8\pi m_D m_{D^*}}{m_{DD^*}^2 a}\, \frac{1}{E-(m_D + m_{D^*} - E_b)}.
\end{align}
The residue is proportional to  the square of the amplitude 
for $\bar{D}^0D^{*0} \rightarrow X$:
\begin{align}
 {\cal A}[\bar{D}^0D^{*0} \rightarrow X]
 =
\left(\frac{16\pi M_X m_D m_{D^*}}{m_{DD^*}^2 a}\right)^{1/2}.
\end{align}

Our strategy is to manipulate the decay rate into a form that
includes as a factor the differential decay rate for $\Upsilon\to B^+B^-$.
The first step is to integrate over the component $\ell_0$ of the loop momentum.
The dominant contribution to the integral over $\ell_0$ 
in (\ref{eq:amplitude})   
comes from the particle poles in the 
propagators for the $\bar{D}^0$ and $D^{*0}$ mesons:
\begin{align}
\int \frac{d\ell_0}{2\pi} \,
\frac{1}{(q+\ell)^2-m_D^2+i\epsilon} \,
\frac{1}{(q'-\ell)^2-m_{D^*}^2+ i\epsilon}
= \frac{i}{4 E_D E_{D^*} ( E_D + E_{D^*}-Q_0 )}.
\label{eq:denominator}
\end{align}
In the rest frame of $X$, the energies are 
$E_D=(m_D^2+\bm{\ell}^2)^{1/2}$, 
$E_{D^*}=(m_{D^*}^2+\bm{\ell}^2)^{1/2}$ and $Q_0=M_X$.
Expanding the denominator to lowest order in $\ell$ and $1/a$, 
(\ref{eq:denominator}) becomes 
\begin{align}
\frac{1}{4 E_D E_{D^*} ( E_D + E_{D^*}-Q_0 )} \simeq 
\frac{m_{DD^*}}{2 m_D m_{D^*}} \frac{1}{\bm{\ell}^2 + 1/a^2},
\end{align}
which is proportional to the momentum-space wavefunction 
$\psi(\ell)$ in (\ref{psi}).

If not for the loop momenta,
the product of the $B^+$ propagator in (\ref{eq:amplitude}) and
its complex conjugate could be expressed as a Breit-Wigner resonance
factor. If we take into account the loop momenta, that product can be 
approximated as 
\begin{align}
&\frac{i}{(p+\ell)^2-m_B^2+ im_B\Gamma_B }
\left( \frac{i}{(p+\ell')^2-m_B^2+ im_B\Gamma_B }\right)^* 
\nonumber \\
&\simeq \frac{1}{p\cdot(\ell-\ell') + (\ell^2-\ell'^2)/2 + im_B\Gamma_B} \, 
\frac{im_B\Gamma_B}{(p^2-m_B^2)^2+(m_B\Gamma_B)^2}
\nonumber \\
&\simeq \frac{1}{
k \cdot(\ell-\ell')+im_B\Gamma_B}
~i\pi \delta(p^2-m_B^2).
\label{eq:propagator}
\end{align}
In the second line, we expressed the product of propagators in terms of a
difference between propagators and took the limit $\ell\to 0$ and
$\ell'\to 0$ in that difference to get a Breit-Wigner resonance factor.
In the third line, we took the limit $\Gamma_B\to 0$ in the resonance
factor to get a delta function.
We also used
the relations $\ell^2 = -2 q\cdot\ell$ and 
$ \ell'^2=-2q\cdot\ell'$,
which follow from the fact that $q$, $q+\ell$ and $q+\ell'$
are all on the mass-shell of the $\bar{D}^0$ meson.
The inner product $k\cdot (\ell-\ell')$ in the denominator of
(\ref{eq:propagator}) can be expanded in powers of the momenta
$\bm{\ell}$ and $\bm{\ell}'$. The terms $q\cdot \ell$ and $q\cdot\ell'$
are already second order and could be neglected. However, for the
purpose of evaluating the integral over $\bm{\ell}$, it is more
convenient to use the fact the $q\cdot\ell$ and $q\cdot\ell'$ are second
order to replace $k^\mu$ by a vector $k_Q^\mu$ whose  
$\mu=0$ component vanishes in
the rest frame of $X$:
\begin{align}
k_Q^{\mu}\equiv k^{\mu}-\frac{k \cdot Q}{Q^2} Q^{\mu}.
\label{eq:k_Q}
\end{align}
The expression for $q$ in (\ref{eq:k_Q}) implies $k_Q^\mu=p_Q^\mu$. Thus
our approximation for the inner product in the denominator of
(\ref{eq:propagator}) can be written
\begin{align}
 k\cdot(\ell-\ell') 
 &\simeq
 p_Q \cdot (\ell-\ell').
\end{align}

We can integrate in the momentum $p$ of the $B^+$ using the
identity
\begin{align}
 1=\int \frac{dp^2}{2\pi} \int \frac{d^3p}{(2\pi)^3 2p_0} (2\pi)^4\delta^{(4)}
(p-k-q).
\end{align}
The integral over $p^2$ can be evaluated using the delta function
in (\ref{eq:propagator}). After similar manipulations involving the
$B^-$ momentum, our decay rate through the first pathway 
can be written
\begin{align}
\Gamma_1[\Upsilon  \rightarrow X h h'] &=
\int d\Gamma[\Upsilon \rightarrow B^+B^-]
\left|{\cal A}[B^+ \rightarrow \bar{D}^0 h]\right|^2
\left|{\cal A}[B^- \rightarrow D^{*0} h']\right|^2
\nonumber \\
&\times \int
(2\pi)^4 \delta^{(4)}(p-q-k) \frac{d^3k}{(2\pi)^3 2 k_0} \,
(2\pi)^4 \delta^{(4)}(p'-q'-k') \frac{d^3k'}{(2\pi)^3 2 k'_0}
\nonumber \\
&\times 
\frac{\pi M_X}{m_D m_{D^*} a}\, 
\int \,I \, 
 \frac{d^3Q}{(2\pi)^3 2 Q_0} \, , 
\label{eq:rate}
\end{align}
where  $d\Gamma[\Upsilon \rightarrow B^+B^-]$ is the differential
decay rate for $\Upsilon$ into $B^+$ and $B^-$ with momenta $p$ and $p'$.
In the rest frame of $X$, 
the factor $I$ in (\ref{eq:rate}) is given by the integral
\begin{align}
I = -\int \frac{d^3\ell}{(2\pi)^3}\frac{d^3\ell'}{(2\pi)^3}\,
\frac{1}{-\bm{p}_Q\cdot (\bm{\ell}-\bm{\ell}')+i \epsilon}\,
\frac{1}{\bm{p'}_Q\cdot(\bm{\ell}-\bm{\ell}')+i\epsilon}\,
\frac{1}{\bm{\ell}^2+1/a^2}\,
\frac{1}{\bm{\ell}'^2+1/a^2}.
\label{eq:I1}
\end{align}
We have replaced the terms $im_B\Gamma_B$ in the propagators by 
$i\epsilon$, because the integral has a well-behaved limit as 
$\Gamma_B\to 0$. 

In order to evaluate $I$,
we first combine both the denominators $\bm{\ell}^2+m^2$ and
$\bm{\ell}'^2+m^2$ where $m=1/a$ 
and the denominators
$-\bm{p}_Q\cdot(\bm{\ell}-\bm{\ell}')+i\epsilon$
and $\bm{p}'_Q\cdot(\bm{\ell}-\bm{\ell}')+i\epsilon$ into squared 
denominators using Feynman parameters $x$ and $z$. We then combine the
squared denominators using an integral over $y$:
\begin{align}
&\frac{1}{-\bm{p}_Q\cdot (\bm{\ell}-\bm{\ell}')+i \epsilon}\,
\frac{1}{\bm{p}'_Q\cdot(\bm{\ell}-\bm{\ell}')+i\epsilon}\,
\frac{1}{\bm{\ell}^2+m^2}\,
\frac{1}{\bm{\ell}'^2+m^2} 
\nonumber \\
&= \int_0^1 dx \,
\frac{1}{[x\bm{\ell}^2+(1-x)\bm{\ell}'^2+m^2]^2}
\int_0^1 dz \,
\frac{1}{[\bm{C}(z)\cdot (\bm{\ell}-\bm{\ell}')+i\epsilon]^2} 
\nonumber \\
&= 6\int_0^1dx\int_0^1dz 
\int_0^{\infty} dy \, \frac{y}{[x\bm{\ell}^2+(1-x)\bm{\ell}'^2+m^2
+ y \bm{C}(z)\cdot (\bm{\ell}-\bm{\ell}')+ i\epsilon ]^4},
\end{align}
where $\bm{C}(z)= -z\bm{p}_Q +(1-z) \bm{p}'_Q$.
The integrals over the momenta $\bm{\ell}$ and
$\bm{\ell'}$ can be simplified by first shifting them
and then rescaling them by factors of $x$ and $1-x$:
\begin{align}
\int & \frac{d^3\ell}{(2\pi)^3}\frac{d^3\ell'}{(2\pi)^3}
\frac{1}{[x\bm{\ell}^2+(1-x)\bm{\ell}'^2+m^2
+ y \bm{C}\cdot (\bm{\ell}-\bm{\ell}')+ i\epsilon ]^4}
 \nonumber \\
&= \int \frac{d^3\ell}{(2\pi)^3}\frac{d^3\ell'}{(2\pi)^3}
\frac{1}{[ x\bm{\ell}^2 + (1-x)\bm{\ell}'^2+m^2 
- y^2\bm{C}^2/(4x(1-x))+i\epsilon]}
\nonumber \\
&= x^{-3/2} (1-x)^{-3/2} 
 \int \frac{d^3\ell}{(2\pi)^3}\frac{d^3\ell'}{(2\pi)^3}
 \frac{1}{[\bm{\ell}^2+\bm{\ell}'^2+m^2
- y^2\bm{C}^2/(4x(1-x)) +i\epsilon]^4}.
\end{align}
The integrals over $y$ and then $x$ can now be evaluated analytically. The
resulting expression for $I$ is 
\begin{align}
 I = 4\pi \int_0^1 dz \frac{1}{\bm{C}(z)^2 }
\int \frac{d^3\ell}{(2\pi)^3}\frac{d^3\ell'}{(2\pi)^3}
\frac{1}{[\bm{\ell}^2+\bm{\ell}'^2+1/a^2]^3}.
\label{eq:I2}
\end{align}
The integral over $\ell$ and $\ell'$ has a logarithmic 
ultraviolet divergence. It can be regularized by subtracting the
integral with $m=1/a$ replaced by an ultraviolet cutoff 
$\Lambda \gg m$: 
\begin{align}
\int \frac{d^3\ell}{(2\pi)^3}\frac{d^3\ell'}{(2\pi)^3}
\frac{1}{[\bm{\ell}^2+\bm{\ell}'^2+m^2]^3}
-\left(m\to\Lambda\right)
= \frac{1}{64\pi^3} \log \frac{\Lambda}{m}.
\end{align}
The appropriate choice for the cutoff is 
$\Lambda=m_\pi$, since the region of validity of the expression
(\ref{psi}) for the wavefunction of $X$ is $|\bm{k}|\ll m_\pi$.
The integral over $z$ in (\ref{eq:I2}) can be expressed 
in a manifestly Lorentz-invariant form. 
Our final expression for the integral $I$ in (\ref{eq:I1}) is  
\begin{align}
I = \frac{1}{16\pi^2}\log (a m_\pi) \, 
  J(p_Q, p'_Q),
\label{eq:I}
\end{align}
where $J(p_Q,p'_Q)$ is 
a Lorentz-invariant function of $p_Q$ and $p'_Q$ 
defined by the integral
\begin{align}
 J(p_Q, p'_Q)=\int_0^1 dz 
\frac{-1}{\left[z p_Q -(1-z) p'_Q \right]^2}. 
\end{align}
The integral can be evaluated analytically:
\begin{align}
J(p_Q, p'_Q)
=  \frac{1}{D(p_Q, p'_Q)} 
\left( 
  \arctan \frac{-P_Q\cdot p_Q }{D(p_Q, p'_Q)}
 +\arctan \frac{-P_Q\cdot p'_Q}{D(p_Q, p'_Q)}
\right),
\label{eq:J}
\end{align}
where 
$p_Q$, $p'_Q$ and $P_Q$ are all defined by (\ref{eq:k_Q}) and 
$D(p_Q, p'_Q) =\left[p_Q^2 p_Q'^2-(p_Q\cdot p'_Q)^2\right]^{1/2}$.
In the rest frame of $X$ where $p_Q$ and $p'_Q$ are spacelike,
$D(p_Q, p'_Q)=|\bm{p}_Q\times \bm{p}'_Q|$.

It remains only to evaluate the integrals over $k$, $k'$ and $Q$ in (\ref{eq:rate}). 
The integral over $k$ can be evaluated by 
using the identity 
\begin{align}
 \int \frac{d^3k}{(2\pi)^3 2k_0}(2\pi)^4\delta^{(4)}(p-q-k)
 = 2\pi\delta((p-q)^2-m_h^2),
\end{align}
and similarly for $k'$.
The two remaining delta functions can then be used to evaluate the integral
over $Q$:  
\begin{align}
\int \frac{d^3Q}{(2\pi)^3 2Q_0}
2\pi \delta((p-q)^2 - m_h^2)
\, 2\pi\delta((p'-q')^2-m_h'^2) 
&= \frac{M_X^2}
 { 4m_D m_{D^*} M_\Upsilon (M_\Upsilon^2-4m_B^2)^{1/2}}.
\label{eq:delta}
\end{align}
Our final expression for the decay rate through the first pathway is
\begin{align}
\Gamma_{1} [\Upsilon\to Xhh']
=&\Gamma[\Upsilon\to B^+B^-]
\, \left|{\cal A} [B^+\to \bar{D}^0 {h} ]\right|^2 
\, \left|{\cal A} [B^-\to D^{*0} {h}' ]\right|^2
\nonumber \\
& \hspace{2.5cm} 
\times 
\frac{M_X J(p_Q, p'_Q)}{64\pi m_{DD^*} M_\Upsilon (M_\Upsilon^2 -4m_B^2)^{1/2}}\,
\frac{\log (am_\pi)}{am_{DD^*}}.
\label{eq:finalrate}
\end{align}
The function $J(p_Q, p'_Q)$ given explicitly in (\ref{eq:J})
is a Lorentz-invariant scalar function of the momenta
$p$, $p'$ and $Q$.
The Lorentz invariants are $p^2=p'^2=m_B^2$, $Q^2=M_X^2$ and 
\begin{align*}
  \begin{split}
   p\cdot Q &= \frac{M_X}{2m_D}\left(m_B^2+m_D^2-m_h^2 \right),
\\
   p'\cdot Q &= \frac{M_X}{2m_{D^*}}\left(m_B^2+m_{D^*}^2-m_{h'}^2 \right),
\\
   p\cdot p' &= \frac{1}{2}\left( M_{\Upsilon}^2 - 2 m_B^2 \right).
  \end{split}
\end{align*}
The decay rate through the second pathway is obtained by replacing 
$\bar D^0$ and $D^{*0}$ by $\bar D^{*0}$ and $D^{0}$.
The inner products between the momenta $k$, $k'$ and $Q$ of the
final-state particles are completely determined by the hadron masses.
Since these inner products are different for the two pathways, they
produce distinct momentum configurations and they therefore cannot interfere.

The rate for the decay $\Upsilon\to B^+B^-$ followed by the decays 
$B^+\to \bar{D}^0 h $ and $B^-\to D^{*0} h'$ is 
\begin{align}
 \Gamma_1[\Upsilon\to \bar{D}^0D^{*0} hh']
 = \Gamma[\Upsilon\to B^+B^-] \, 
   {\cal B}[B^+\to \bar{D}^0 h] \,
   {\cal B}[B^-\to D^{*0} h'].
\label{eq:DDhh}
\end{align}
The expression for the branching fraction ${\cal B}[B^+\to \bar{D}^{0} h]$ is given
in (\ref{eq:B}). The amplitude 
${\cal A}[B^+\to \bar D^0 h]$  in (\ref{eq:finalrate})
can be eliminated in favor of ${\cal B}[B^+\to \bar D^0 h]$.


\end{document}